\documentclass[twocolumn,showpacs,preprintnumbers,amsmath,amssymb]{revtex4}
\usepackage{graphicx}
\usepackage{dcolumn}
\usepackage{bm}

\begin{document}

\preprint{APS/123-QED}
\title{A comparative analysis of tunneling time concepts: Where do transmitted particles
start from, on the average?}

\author{N. L. Chuprikov}
\email{chnl@tspu.edu.ru}
\affiliation{%
Tomsk State Pedagogical University, 634041, Tomsk, Russia}
\altaffiliation[Also at
]{Physics Department, Tomsk State University.}

\date{\today}

\begin{abstract}

In this paper we compare the concept of the tunneling time introduced in
quant-ph/0405028 with those of the phase and dwell times. As is shown, unlike the
latter our definition of the transmission time coincides, in the limit of weak
scattering potentials, with that for a free particle. This is valid for all values of
the particle's momentum, including the case of however slow particles. All three
times are also considered for a resonant tunneling. In all the cases the main feature
to distinguish our concept from others is that the average starting point of
transmitted (reflected) particles does not coincide with that of all particles. One
has to stress here that there is no such an experiment which would give coordinates
of all the three points, simultaneously. For measuring the position of the average
starting point of transmitted particles we propose an experimental scheme based on the
Larmor precession effect.
\end{abstract}
\maketitle

\newcommand {\uta} {\tau_{tr}}
\newcommand {\utb} {\tau_{ref}}
\newcommand{\ppp}{\mbox{\hspace{5mm}}}
\newcommand{\ooo}{\mbox{\hspace{3mm}}}
\newcommand{\ooa}{\mbox{\hspace{1mm}}}

\section{Introduction}

As is known \cite{Ha2,La1,Olk,Ste,Mu0,Nu0}, the question of the time spent by a
quantum particle in the barrier region is still controversial. At present there are
many different definitions of this quantity, however none of them is commonly
accepted. At the same time, for a given potential and initial state of a particle,
i.e., in the standard setting of the tunneling problem, the above question is evident
to imply an unique answer.

In our previous paper \cite{Ch5} we have introduced the concepts of the transmission
and reflection times based on a separate description of the transmission and
reflection processes. By our approach, tunneling is a combined random process to
consist from two alternative elementary ones, transmission and reflection. For a given
potential and initial state of a particle, we have found an unique pair of solutions
to the Schr\"odinger equation, which describe separately these processes. This
permits one to follow the centers of "mass" (CMs) of the transmitted and reflected
wave packets, at all stages of scattering, and, as a consequence, to calculate the
times spent by the CMs in the barrier region. These characteristic times are treated
in our formalism as the (average) transmission and reflection times for a particle.

Note, unlike the standard wave-packet analysis (SWPA) our approach predicts that the
average starting point of transmitted (and reflected) particles does not coincide with
the starting point of the CM of the incident wave packet to describe all particles. By
this reason, our transmission and reflection times differ essentially from the
corresponding phase times derived in the SWPA.

We have to stress that our approach agrees entirely with the foundations of quantum
theory, for the wave functions for transmission and reflection to underlay it are
solutions to the Schr\"odinger equation. Besides, one has to bear in mind that the
average starting point of transmitted particles and that of all scattering particles
cannot be measured simultaneously, in principle.

In this paper we show that the formal comparison of the well-known phase and dwell
times with ours speaks in favor of the latter. And, what is more important, our
formalism can be experimentally verified. Due to the Larmor precession of the
particle's spin in an external magnetic field applied beyond the barrier region, one
can measure, in principle, the average starting point of transmitted (or reflected)
particles.

Note, apart from the phase time, the dwell and Larmor time are, perhaps, the most
cited concepts of the tunneling time. Thus, with taking into account that the Larmor
time coincides with the dwell time, it is useful to compare our concept of the
tunneling time with those of the phase and dwell times.

\section{Critique of the dwell-time concept}

In \cite{Ch5} we pointed to the principal shortcomings of the phase time concept. Now,
before a formal comparing of the three concepts, we want to dwell on the principal
shortcomings of the dwell time concept.

We begin with the fact that calculating the average values of the particle's position
and momentum makes sense only when this calculation is performed separately for the
subensembles of transmitted and reflected particles. As regards the average values of
these quantities calculated over the whole ensemble of particles, they behave
non-causally in the course of the scattering process and hence cannot be interpreted
as the expectation values of the position and momentum of a scattering particle.

By our approach, introducing an observable to describe tunneling, without
distinguishing transmission and reflection, is meaningless. Quantum mechanics does not
imply computing the expectation values of physical quantities for a tunneling
particle, which would be common for transmission and reflection. Figuratively
speaking, it is merely impossible to "pack up" the properties of the alternative
processes into one common characteristics. This concerns entirely the dwell time.

Thus, by our approach, introducing this time scale is questionable from the most
beginning. Besides, it is evident that a proper definition of the transmission time
for any spatial interval should be valid irrespective of the displacement of this
interval on the $OX$-axis. However, the definition of the dwell time is evident to
violate this requirement. Indeed, as is known, this time is defined as the ratio of
the probability to find a particle within the barrier region to the incident flux. We
consider this definition as a purely speculative one. For the used here normalization
by the incident flux has no solid physical basis.

Let us assume, for example, that we study the motion of a particle in the spatial
region of the same width but shifted, with respect to the barrier region, toward the
transmission domain. What flux should be used in defining the dwell time for this
region? Of course, if this spatial region lies entirely in the transmission region,
then it is naturally to use for this purpose the transmitted flux (and, as a
consequence, a resulting dwell time will describe transmitted particles only).
However, if the shifted spatial region coincides partly with the barrier region, then
neither incident nor transmitted flux cannot be used in the above definition.

Consider another example. Let the spatial region investigated be the right half of the
localization region of the rectangular potential barrier. At the first glance, the
dwell time for this case should be the ratio of the probability to find a particle
within this region to the incident flux. However, by our approach, if the particle
impinges the barrier from the left, then in the case of reflection it never enters the
right half of a symmetrical potential barrier. It is evident to be meaningless to
take into account the reflected part of the quantum ensemble of particles, in timing a
particle in this spatial region.

\section{Rectangular potential barriers}

Now we proceed to a formal comparison of the properties of our time scale with those
of the phase and dwell times. Note firstly that in any approach the tunneling time for
the particle with a given value of $k$ can be written as $m D_{eff}(k)/\hbar k$. In
this paper we analyze the behaviour of $D_{eff}(k)$, in the case of rectangular
potential barriers, for the phase ($\tau_{phase}$), dwell ($\tau_{dwell}$) and
tunneling time ($\tau_{tun}$) introduced in our previous paper \cite{Ch5}. Of course,
unlike $\tau_{phase}$ and $\tau_{tun}$, $\tau_{dwell}$ describes all particle of a
quantum ensemble. However, one has to bear in mind that in the case of rectangular
potential barriers the tunneling times for transmission and reflection are equal. It
is naturally to expect that in this case the dwell time should coincide by value with
the transmission time.

Taking into account the expression for $\tau_{phase}$, $\tau_{dwell}$ (see, for
example, \cite{But}) and $\tau_{tun}$, one can easily to obtain the expressions for
the corresponding effective barrier's widths. Let
\[\tau_{phase}(k)\equiv \frac{m}{\hbar k} D_{phase}(k), \ooo \tau_{dwell}(k)\equiv
\frac{m}{\hbar k} D_{dwell}(k),\] and
\[\tau_{tun}(k)\equiv \frac{m}{\hbar k} d_{eff}(k).\] For the below-barrier case
($E\le V_0$) we have

\[D_{phase}=\frac{1}{\kappa}\cdot \frac{2\kappa d k^2(\kappa^2-k^2)+\kappa_0^4\sinh(2\kappa
d)} {4k^2\kappa^2+ \kappa_0^4\sinh^2(\kappa d)},\]
\[D_{dwell}=\frac{k^2}{\kappa}\cdot \frac{2\kappa d (\kappa^2-k^2)+\kappa_0^2\sinh(2\kappa
d)} {4k^2\kappa^2+ \kappa_0^4\sinh^2(\kappa d)},\]
\[d_{eff}=\frac{4}{\kappa}\cdot \frac{\left[k^2+\kappa_0^2\sinh^2\left(\kappa
d/2\right)\right] \left[\kappa_0^2\sinh(\kappa d)-k^2 \kappa d\right]} {4k^2\kappa^2+
\kappa_0^4\sinh^2(\kappa d)}\]
where $\kappa=\sqrt{2m(V_0-E)/\hbar^2}.$

For the above-barrier case ($E\ge V_0)$ ---
\[D_{phase}=\frac{1}{\kappa}\cdot \frac{2\kappa d k^2(\kappa^2+k^2)-\kappa_0^4\sin(2\kappa
d)} {4k^2\kappa^2+ \kappa_0^4\sin^2(\kappa d)},\]
\[D_{dwell}=\frac{k^2}{\kappa}\cdot \frac{2\kappa d (\kappa^2+k^2)-\kappa_0^2\sin(2\kappa
d)} {4k^2\kappa^2+ \kappa_0^4\sin^2(\kappa d)},\]
\[d_{eff}=\frac{4}{\kappa}\cdot \frac{\left[k^2-\beta
\kappa_0^2\sin^2\left(\kappa d/2\right)\right]\left[k^2 \kappa d-\beta
\kappa_0^2\sin(\kappa d)\right]} {4k^2\kappa^2+\kappa_0^4\sin^2(\kappa d)}\] where
$\kappa=\sqrt{2m(E-V_0)/\hbar^2};$ $\beta=1$ if $V_0>0$, otherwise, $\beta=-1$. In
both the cases $\kappa_0=\sqrt{2m|V_0|/\hbar^2}$.

In \cite{Ch5} we treated the case when the potential barrier is localized in the
region $[a,b]$, and the CM of the incident wave packet to describe all particles is,
at $t=0$, at the point $x=0$; $a>0$. Remind that in accordance with the standard
wave-packet analysis namely this spatial point should be considered as the average
starting point both for transmitted and for reflected particles. However, by our
approach, this is not the case. As was shown in \cite{Ch5}, the average starting
points, $x_{start}^{tr}$ and $x_{start}^{ref}$, of transmitted and reflected
particles, respectively, do not coincide with that for all particles.

For the above initial condition, in the case of symmetrical potential barriers, we
have $x_{start}^{tr}(k)=x_{start}^{ref}(k)=x_{start}(k)$ where
\[x_{start}(k)= -2\frac{\kappa_0^2}{\kappa}
\frac{(\kappa^2-k^2)\sinh(\kappa d)+k^2 \kappa d \cosh(\kappa d)} {4k^2\kappa^2+
\kappa_0^4\sinh^2(\kappa d)},\]
\[x_{start}(k)= -2\beta \frac{\kappa_0^2}{\kappa} \cdot
\frac{(\kappa^2+k^2)\sin(\kappa d)-k^2 \kappa d \cos(\kappa d)} {4k^2\kappa^2+
\kappa_0^4\sin^2(\kappa d)},\] for $E<V_0$ and $E\ge V_0$, respectively. Note, these
quantities are such that (see \cite{Ch5})
\begin{eqnarray} \label{100}
D_{phase}(k)=d_{eff}(k)-x_{start}(k),
\end{eqnarray}
from which a formal connection between the phase-time concept and ours is seen
explicitly.

\section{Characteristic times in the limit of weak scattering potentials}

As is known, by the standard in quantum mechanics timing procedure (based on timing
the CM of a wave packet), for a free particle with the well defined momentum $\hbar
k$ the average time spent by the particle in the spatial region of width $d$ is equal
to $\tau_{free}(k;d)$ where $\tau_{free}(k;d)=m d/\hbar k$; $m$ is the particle's
mass, $k=\sqrt{2mE/\hbar^{2}}$, $E$ is the particle's energy. This expression is
valid for any value of $k$. In particular, in the limit $k\to 0$, $\tau_{free}(k;d)$
diverges as $k^{-1}$.

It is obvious that when the potential energy of a particle in the barrier region
diminishes (this case is named here as the limit of weak scattering potentials) then
$D_{eff}(k)$ should approach $d$, for any value of $k$. In the limit of weak
scattering potentials the $k$-dependence of a true tunneling time must approach the
function $\tau_{free}(k;d)$. In particular, for infinitesimal potentials the
transmission time should diverge as $k^{-1}$ when $k\to 0$. This requirement should
be considered as a touchstone in solving the tunneling time problem. For example, in
the case of the rectangular barrier of width $d$ and height $V_0$ this should take
place when $V_0 d\to 0$.

At the first glance, all three times behave properly in the limit of weak scattering
potentials. Indeed, in all the cases, if $k\neq 0,$ $D_{eff}(k)/d\to 1$ when
$\kappa_0d\to 0$. But, as it was stressed above, this should be valid also for
however small values of $k$. As will be seen from the following, only $\tau_{tun}$
obeys this requirement.

One can easily show that for $\kappa_0d\neq 0$ and $k=0$
\begin{eqnarray} \label{1}
\frac{D_{phase}}{d}=\frac{2}{\kappa_0 d\tanh(\kappa_0 d)}, \ppp \frac{D_{dwell}}{d}=0,
\end{eqnarray}
\begin{eqnarray} \label{2}
\frac{d_{eff}}{d}=\frac{2}{\kappa_0d}\tanh\left(\frac{\kappa_0 d}{2}\right),\ooa
\frac{x_{start}}{d}=-\frac{2}{\kappa_0d \sinh(\kappa_0 d)}
\end{eqnarray}
for $V_0>0$; for $V_0<0$
\begin{eqnarray} \label{3}
\frac{D_{phase}}{d}=-\frac{2}{\kappa_0d\tan(\kappa_0 d)}, \ppp \frac{D_{dwell}}{d}=0,
\end{eqnarray}
\begin{eqnarray} \label{4}
\frac{d_{eff}}{d}=\frac{2}{\kappa_0d}\tan\left(\frac{\kappa_0 d}{2}\right),\ppp
\frac{x_{start}}{d}=\frac{2}{\kappa_0 d\sin(\kappa_0 d)}.
\end{eqnarray}

Note, in the long-wave limit, $D_{dwell}/d=0$ irrespective of $\kappa_0d$. The
corresponding limits for the phase time and $\tau_{tun}$ depend on $\kappa_0d$.
Moreover, for $k=0$, as $\kappa_0d\to 0$, $D_{phase}/d\to \infty$ for barriers
($V_0>0$), and $D_{phase}/d\to -\infty$ for wells ($V_0<0$). As regards our
definition, in this limit $d_{eff}/d=1$ both for barriers and wells.

Figs. 1-4 show the dependence of $D_{eff}/d$ on $E/V_0$, for all three characteristic
times, for weak scattering potentials. Figs. 1, 2 correspond to the narrow ($d=0.5
nm$) rectangular barrier ($V_0=0.25 eV$) and well ($V_0=-0.25 eV$), respectively.
Figs. 3, 4 correspond to the wide ($d=50 nm$) rectangular barrier ($V_0=0.00025 eV$)
and well ($V_0=-0.00025 eV$), respectively. In all cases $m=0.067m_e$ where $m_e$ is
the mass of an electron.

So, in the limit of weak scattering potentials $d_{eff}/d=1$ for all values of $k$.
That is, our definition of the tunneling time guarantees the passage to the free
particle case, in this limit. However, this is not the case for the phase and dwell
times. By these concepts, in the limit of weak scattering potentials the time spent
by a slow particle in the barrier region differ essentially from $\tau_{free}(k;d)$.
In particular, in comparison with $\tau_{free}$ the dwell-time concept predicts
anomalously short times spent by a slow particle in the barrier region.

To explain this fact, let us remember once more that the dwell time was introduced as
the ratio of the probability to find a particle in the barrier region to the incident
probability flux. It is clear that for $\kappa_0\neq 0$, in the long-wave limit, the
number of particles in the barrier region is proportional to $k^2$. That is, a
particle with a however small value of $k$ does not enter the barrier. Taking also
into account the fact that the incident flux $\sim k$ in this limit, we obtain
$\tau_{dwell}\sim k$ instead of $\tau_{dwell}\sim k^{-1}$.

This property of the dwell time is kept for a however small value of $\kappa_0d$.
That is, strictly speaking, the dwell-time concept does not imply the passage to the
case of a free particle, in the limit of weak scattering potentials. This fact
evidences that the dwell time is ill-defined and cannot serve as the characteristic
time to describe the dynamical properties of a tunneling particle.

As is seen from the figures, the phase-time concept does not guarantee the above
passage, too. For slow particles, $|D_{phase}|/d$ may be however large, being negative
by value in the case of wells.

Note, to compare the properties of the phase times and ours is of a particular
importance. "Where do transmitted particles start from, on the average?" is the main
intriguing question to arise in this case. Remind, in contrast with the phase-time
concept to imply that in the above setting the tunneling problem transmitted
particles start, on the average, from the point $x=0$, our formalism says that this
point is $x_{start}$.

Exps. (\ref{1}) and (\ref{3}) show that in the limit of weak scattering potentials the
phase-time concept predicts an abnormal divergence of the transmission time at $k\to
0$. We have to stress that such a behaviour of the phase time takes place even if the
transmission coefficient, $T$, approaches unit. Indeed, let us consider the case when
$\kappa_0\neq 0$ and $k=d/\lambda^2\to 0$; $\lambda$ is fixed. One can easily show
that in this case \[T=\frac{4}{4+\lambda^4\kappa_0^4},\] that is, $T$ is constant in
this limit. For $V_0>0$ we have
\[\frac{D_{dwell}}{d}=\frac{4}{4+\lambda^4\kappa_0^4},\ppp
\frac{D_{phase}}{d}=\frac{2\lambda^4
\kappa_0^2}{4+\lambda^4\kappa_0^4}\cdot\frac{1}{d^2};\] for $V_0<0,$
\[\frac{D_{dwell}}{d}=0,\ppp
\frac{D_{phase}}{d}=-\frac{2\lambda^4
\kappa_0^2}{4+\lambda^4\kappa_0^4}\cdot\frac{1}{d^2}.\] In both the cases
\[\frac{d_{eff}}{d}=1, \ppp x_{start}=-D_{phase}.\]
As is seen, if $\lambda^4\kappa_0^4\ll 4$, the barrier is transparent in this limit.
What is more important, for wells $D_{dwell}=0$ even when $T\approx 1$. As regards
$D_{phase}$, it diverges in this limit, both for barriers and wells. This means, in
turn, that in this case $x_{start}$ in our approach diverges too (of course, for
finite wave packets the average value of $|x_{start}|$ does not exceed the
wave-packet's half-width).

Figs. 5-8 show the time evolution of the wave packets $\Psi_{full}$, $\Psi_{tr}$ and
$\Psi_{ref}$ to describe all, transmitted and reflected particles, respectively. The
potential barrier considered is a rectangular well: $V_0=-712eV$, $d=1.08\times
10^{-5}nm$; in all cases $m=0.067m_e$. At $t=0$ (see Fig. 5) the state of a particle
is described by the Gaussian wave packet whose half-width equals $15 nm$, the average
particle's energy equals $0.00641eV$; $a=70 nm$. In this case the norm,
$\overline{R}$, of $\Psi_{ref}$ equals $6.5\cdot 10^{-3}$.

As is seen from Figs. 6-8, this weak scattering potential unexpectedly strongly
changes the shape of the "full" wave packet. When the peak to correspond, at early
times, to the CM of the incident wave packet arrives at the point lying at some
distance of the barrier, a new peak to correspond to the CM of the transmitted wave
packet simultaneously appears behind the barrier region (see Fig.7).

By the SWPA, the transmission time is negative in this case. This property of
$\Psi_{full}(x)$ is usually interpreted as the evidence of an ultrafast or even
superluminal propagation of a particle through the barrier. However, by our approach,
such interpretation of the wave-packet tunneling is wrong. The above behaviour of the
wave-packet's peaks is inherent to interference maxima. It is evident that the
expectation value of the particle's position cannot behave in such a manner. In our
approach, for any $t$, this value for transmitted particles should be calculated over
the wave packet $\Psi_{tr}$ whose motion near the barrier region is regular.

The above property of the "full" wave packet shows once more that there is no direct
causal link between the transmitted and incident wave packets. By our approach, the
tunneling process is a combined stochastic process consisting from two alternative
ones, transmission and reflection, provided that $\Psi_{full}= \Psi_{tr}+\Psi_{ref}$,
where $\Psi_{tr}$ is the wave function to describe transmission and $\Psi_{ref}$ is
that to describe reflection (see \cite{Ch5}).

It is important to remind here that the norms of $\Psi_{full}$ and $\Psi_{tr}$ are
practically equal in the case investigated. However, due to the interference terms
(which are proportional to $\sqrt{\overline{R}}$, rather than $\overline{R}$) the
difference between the behaviour of the elementary state $\Psi_{tr}$ and combined
state $\Psi_{full}$ (see \cite{Ch5}) is essential. In particular, unlike the "full"
wave packet, the shape of $|\Psi_{tr}(x)|^{2}$ remains practically unaltered during
the scattering event (see Figs. 6-8). The distance ($\approx 2 nm$) between the
average starting points of all and transmitted particles is well larger than the
width of this transparent barrier.

Remind (see \cite{Ch5}), the incoming wave to describe transmitting particles with a
given value of $k$ differs from that for all particles by the factor $\exp(\pm
i\gamma(k))$ where $\gamma(k)=\arctan\sqrt{R/T}$; $R=1-T$. Since
$|x_{start}|=|R^\prime/[2\sqrt{RT}]|$ this shift of the average starting point may be
very large even if the reflection coefficient $R(k)$ is small by value; for the key
role here is played by $R^\prime(k)$; the prime denotes the derivative on $k$. A
similar situation arises in the case of a resonant tunneling.

\section{Resonant tunneling}

One can easily show that the dwell and phase times, distinguished drastically in the
long-wave limit, coincide with each other in the case of a resonant tunneling. Namely,
for $E>V_0$, for $\kappa d=n\pi$ ($n=1,2,\ldots$), where
$k=\sqrt{\beta\kappa_0^2+n^2\pi^2/d^2}$, we have
\[\frac{D_{phase}}{d}=\frac{D_{dwell}}{d}=\frac{\kappa^2+k^2}{2\kappa^2}=
1+\frac{\beta\kappa_0^2d^2}{2n^2\pi^2}.\] At the same time, if $n$ is even, then
\[\frac{d_{eff}}{d}=\frac{k^2}{\kappa^2}=1+\beta\frac{\kappa_0^2d^2}{n^2\pi^2};\]
if $n$ is odd,
\[\frac{d_{eff}}{d}=\frac{k^2-\beta\kappa_0^2}{\kappa^2}\equiv 1;\]
besides,
\[\frac{x_{start}}{d}=(-1)^n\frac{\beta
\kappa_0^2d^2}{2n^2\pi^2}.\] Note, near the resonance point $k_r$ the transmission
coefficient can be written as $T(k)=[1+a_0^2(k-k_r)^2]^{-1}$, where $a_0$ is a length
to characterize the resonance. It is evident that $|x_{start}|=a_0$. So that, the
narrower the resonance peak on $T(k)$, the larger is the value of $|x_{start}|$.

\section{About the possibility of measuring the average starting point of transmitted
particles}

Of course, a new and important result to arise in the framework of the separate
description of transmission and reflection is that the average starting point of
transmitted (and reflected) particles does not coincide with the initial position of
the CM of the incident wave packet to describe the whole ensemble of particles. From
the theoretical point of view, we deal here with three wave fields, $\Psi_{full}$,
$\Psi_{tr}$ and $\Psi_{ref}$, where $\Psi_{full}= \Psi_{tr}+\Psi_{ref}$. Thus, it is
not surprising that due to the interference between $\Psi_{tr}$ and $\Psi_{ref}$ the
starting points of the CMs of the last two wave packets do not coincide, in the
general case, with that of the incident wave packet.

This means, in particular, that the average starting point of transmitted (or
reflected) particles and that of all particles cannot be measured simultaneously. Let
$O_{inc}$ be an observer to study particles before the scattering event. Besides, let
$O_{tr}$ and $O_{ref}$ be observers to study transmitted and reflected particles,
respectively. It is evident that from the viewpoint of $O_{inc}$ particles start, on
the average, from the point $x=0$. This observer cannot separate transmission and
reflection. On the contrary, $O_{tr}$ ($O_{ref}$) cannot measure the average starting
point for all particles, but he can measure the average starting point of transmitted
(reflected) particles. As is follows from our approach, he must find that particles
transmitted by the barrier start, on the average, from the point $x_{start}$.

Of course, in verifying our formalism, measuring the value of $x_{start}$ is of great
importance. As in \cite{But}, we will exploit for this purpose the Larmor precession
of a spin-$1/2$ particle in a small magnetic field. However, our aim is directly
opposite, because timing the scattering particle have already been performed in our
approach.

Let an infinitesimal constant uniform magnetic field $\mathbf{B}$ be applied along the
$z$-axis, everywhere outside the interval $[a-l,b+l]$ on the $x$-axis; we assume that
$a-l>0$, moreover $a-l\gg l_0$ and $l\gg l_0$; $l_0$ is the half-width of the incident
wave packet. When a spin-$1/2$ particle moves in this region, the axis of its
(average) spin will rotate around the $z$-axis with a constant frequency
$\omega_L=g\mu B/\hbar$, where $g$ is the gyromagnetic ratio, $\mu$ is the absolute
value of the magnetic moment of the particle. Of course, we assume that at $t=0$ the
spin axis is not parallel to the $z$-axis. Moreover, we assume also that at this
moment the expectation values $<S_x>$ and $<S_y>$ of the $x$- and $y$-components of
the spin are equal: $<S_x^{(0)}>$ and $<S_y^{(0)}>$.

Let the spin of a particle with the well-defined value of $k$ (the corresponding wave
packet is narrow in $k$-space) be detected at that instant of time, $t_{det}$, when
the CM of the transmitted wave packet arrives at the point $x=b+L$; $L-l\gg l_0$.
Thus, on the average, the particle moves under the magnetic field during the time
$\Delta t_{in}+\Delta t_{out}$, where $\Delta t_{in}=\frac{m}{\hbar
k}(a-l-x_{start})$, $\Delta t_{out}=\frac{m}{\hbar k}(L-l)$. Hence, the expectation
values $<S_x>$ and $<S_y>$ for transmitted particles, at the instant of time
$t=t_{det}$, read as
\[<S_x>=\frac{<S_x^{(0)}>}{\sqrt{2}}\cos\left[\frac{m\omega_L}{\hbar
k}(a+L-2l-x_{start})+\frac{\pi}{4}\right]\]
\[<S_y>=\frac{<S_y^{(0)}>}{\sqrt{2}}\sin\left[\frac{m\omega_L}{\hbar
k}(a+L-2l-x_{start})+\frac{\pi}{4}\right]\]
From this it follows that for this moment in the case of equal values of $<S_x^{(0)}>$
and $<S_y^{(0)}>$ we have
\begin{eqnarray} \label{6}
x_{start}=a+L-2l+\frac{\hbar k}{m\omega_L}\Bigg[\frac{\pi}{4}
-\arctan\left(\frac{<S_y>}{<S_x>}\right)\Bigg]
\end{eqnarray}
Note, according to the SWPA the expression in the righthand side of (\ref{6}) should
be equal to zero.

Thus, Exp. (\ref{6}) can serve as the basis for the experimental checking of our
approach. They provide the way of calculating the value of $x_{start}$ from
experimental data on measuring the particle's spin. We have to stress once more that
these expressions describe the case when an external magnetic field is applied to the
spatial regions where the incident and transmitted wave packets evolve freely.

Note, Exp. (\ref{6}) is valid also for finite wave packets. However, now the
expectation values of $k$ for the incident ($<k>_{inc}$), transmitted ($<k>_{tr}$)
and reflected ($<k>_{ref}$) wave packets are different. Thus, $k$ in (\ref{6}) should
be replaced in this case with $<k>_{tr}$.

\section*{Figure captions}

Fig. 1. $D_{eff}/d$ versus $E/V_0$ for the narrow ($d=0.5 nm$) rectangular barrier
($V_0=0.25 eV$).

\vspace{1cm} \noindent Fig. 2. $D_{eff}/d$ versus $E/V_0$ for the narrow ($d=0.5 nm$)
rectangular well ($V_0=-0.25 eV$).

\vspace{1cm} \noindent Fig. 3. $D_{eff}/d$ versus $E/V_0$ for the wide ($d=50 nm$)
rectangular barrier ($V_0=0.00025 eV$).

\vspace{1cm} \noindent Fig. 4. $D_{eff}/d$ versus $E/V_0$ for the wide ($d=50 nm$)
rectangular well ($V_0=-0.00025 eV$).

\vspace{1cm} \noindent Fig. 5. The $x$-dependence of $|\Psi_{full}|^2$ (solid line)
which represents the Gaussian wave packet with $l_0=15 nm$ and the average kinetic
particle's energy $0.00641 eV$, as well as $|\Psi_{tr}|^2$ (open circles) and
$|\Psi_{ref}|^2$ (dashed line) for the rectangular well ($V_0=-712eV$, $d=1.08\times
10^{-5} nm$, $a=70 nm$); $t=0$.

\vspace{1cm} \noindent Fig. 6. The same as in Fig. 5, but $t=29 ps$.

\vspace{1cm} \noindent Fig. 7. The same as in Fig. 5, but $t=33.5 ps$.

\vspace{1cm} \noindent Fig. 8. The same as in Fig. 5, but $t=38 ps$.

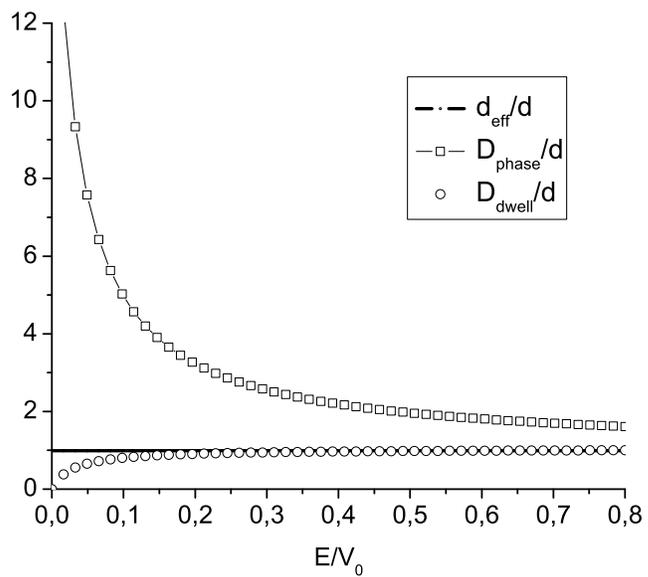
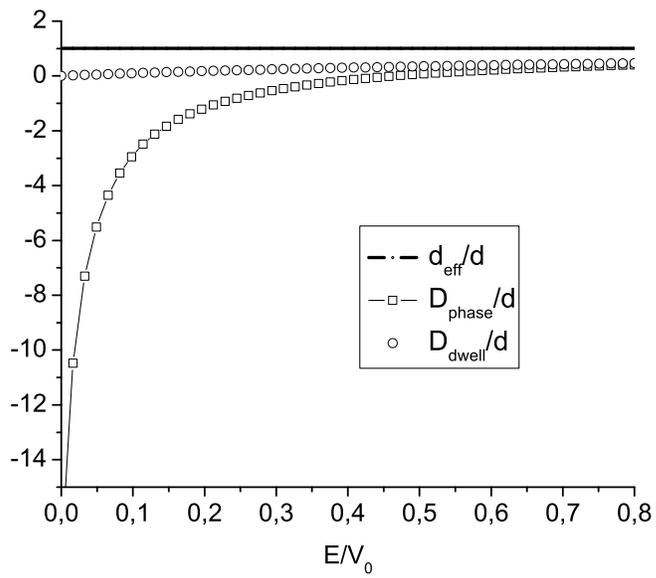
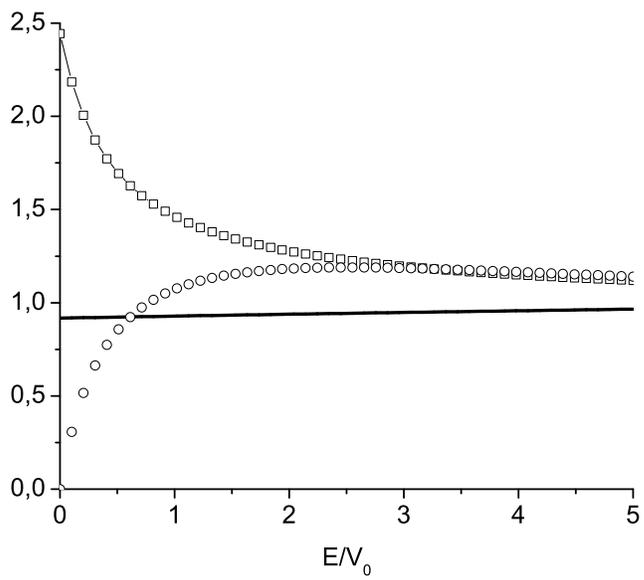
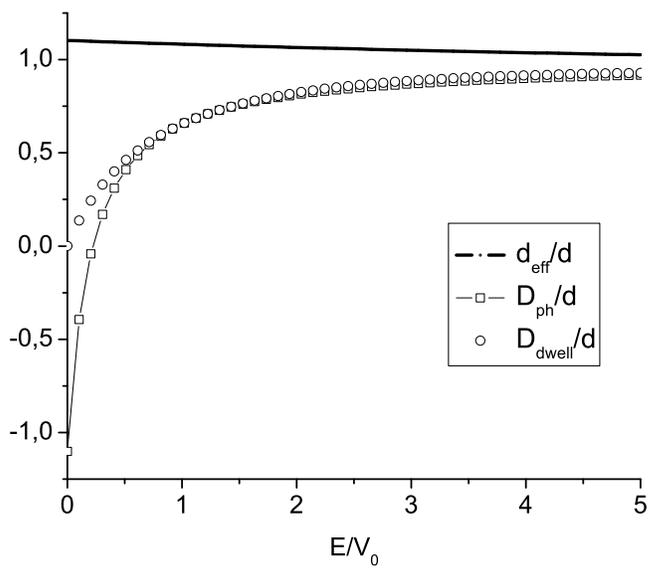
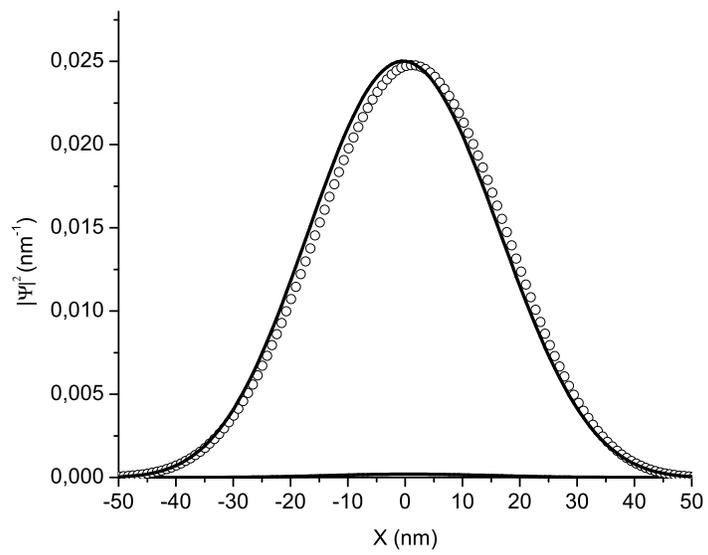
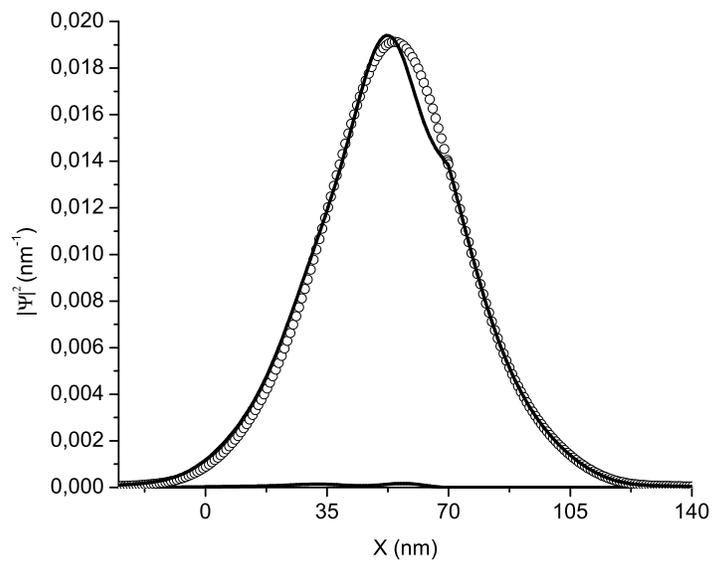
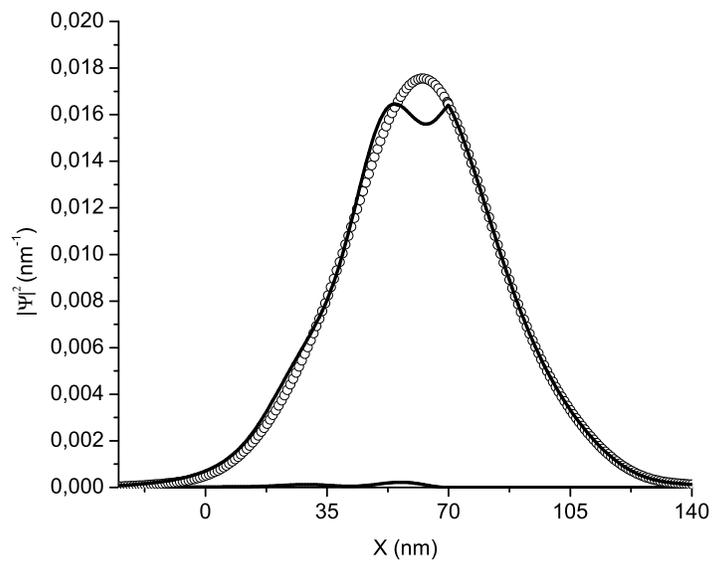
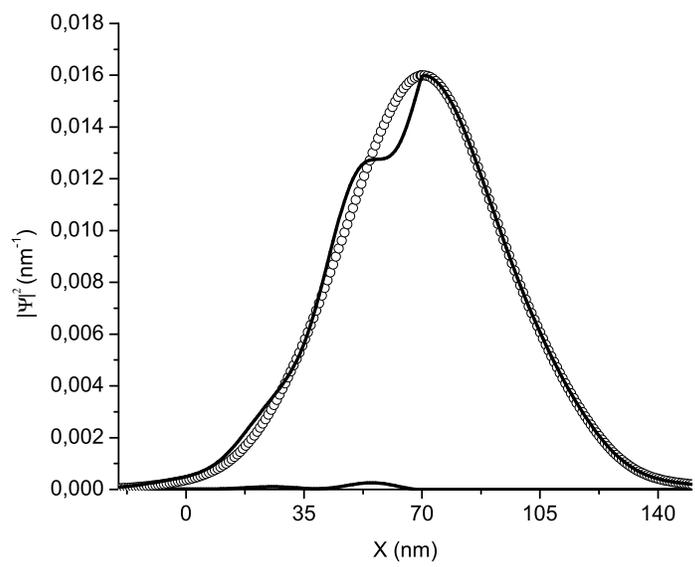
\end{document}